\documentstyle[12pt]{article}

\font\twlgot =eufm10 scaled \magstep1
\font\egtgot =eufm8
\font\sevgot =eufm7
\font\twlmsb =msbm10 scaled \magstep1
\font\egtmsb =msbm8
\font\sevmsb =msbm7

\newfam\gotfam
\def\pgot{\fam\gotfam\twlgot}
\textfont\gotfam\twlgot
\scriptfont\gotfam\egtgot
\scriptscriptfont\gotfam\sevgot
\def\got{\protect\pgot}
\newfam\msbfam
\textfont\msbfam\twlmsb
\scriptfont\msbfam\egtmsb
\scriptscriptfont\msbfam\sevmsb

\def\pBbb{\relax\ifmmode\expandafter\Bb\else\typeout{You cann't use
Bbb in text mode}\fi}
\def\Bb #1{{\fam\msbfam\relax#1}}

\newcommand{\gF}{{\got F}}

\def\thebibliography#1{\bigskip\section*{}\bigskip\list
{$^{\arabic{enumi}}$}{\settowidth\labelwidth{#1}\leftmargin\labelwidth
\advance\leftmargin\labelsep
\usecounter{enumi}}
\def\newblock{\hskip .11em plus .33em minus .07em}
\sloppy\clubpenalty4000\widowpenalty4000
\sfcode`\.=1000\relax}

\let\Large=\large

\def\op#1{\mathop{\fam0 #1}\limits}

\newcommand{\id}{{\rm Id\,}}

\newcommand{\beq}{\begin{equation}}
\newcommand{\eeq}{\end{equation}}
\newcommand{\ben}{\begin{eqnarray}}
\newcommand{\een}{\end{eqnarray}}
\newcommand{\be}{\begin{eqnarray*}}
\newcommand{\ee}{\end{eqnarray*}}
\newcommand{\bea}{\begin{eqalph}}
\newcommand{\eea}{\end{eqalph}}
\newcommand{\cA}{{\cal A}}
\newcommand{\cT}{{\cal T}}

\newcommand{\cD}{{\cal D}}

\newcommand{\cH}{{\cal H}}
\newcommand{\cF}{{\cal F}}

\newcommand{\cO}{{\cal O}}
\newcommand{\bL}{{\bf L}}
\newcommand{\bR}{{\bf R}}
\newcommand{\bC}{{\bf C}}
\newcommand{\bZ}{{\bf Z}}

\newcommand{\bT}{{\bf T}}

\newcommand{\al}{\alpha}

\newcommand{\bt}{\beta}

\newcommand{\la}{\lambda}
\newcommand{\La}{\Lambda}
\newcommand{\f}{\phi}

\newcommand{\Om}{\Omega}
\newcommand{\m}{\mu}

\newcommand{\g}{\gamma}
\newcommand{\G}{\Gamma}

\newcommand{\vt}{\vartheta}
\newcommand{\vf}{\varphi}

\newcommand{\lng}{\langle}
\newcommand{\rng}{\rangle}
\newcommand{\di}{{\rm dim\,}}

\newcommand{\si}{\sigma}
\newcommand{\Si}{\Sigma}
\newcommand{\w}{\wedge}
\newcommand{\wt}{\widetilde}
\newcommand{\wh}{\widehat}
\newcommand{\ol}{\overline}
\newcommand{\dr}{\partial}
\newcommand{\ar}{\op\longrightarrow}
\newcommand{\ot}{\otimes}

\newcommand{\ve}{\varepsilon}

\newcommand{\fl}{\flat}
\newcommand{\sh}{\sharp}

\let\ssection=\section
\renewcommand{\section}{\setcounter{equation}{0}\ssection}

\newcounter{eqalph}
\newcounter{equationa}
\newcounter{remark}
\newcounter{example}
\newcounter{theorem}
\newcounter{proposition}
\newcounter{lemma}
\newcounter{corollary}
\newcounter{definition}
\newenvironment{eqalph}{\stepcounter{equation}
\setcounter{equationa}{\value{equation}}
\setcounter{equation}{0}

\begin{eqnarray}}{\end{eqnarray}\setcounter{equation}{\value{equationa}}}

\def\theremark{\arabic{remark}}
\def\therexample{\arabic{remark}}

\def\thedefinition{\arabic{definition}}

\newenvironment{proof}{\medskip\noindent
{\it Proof:}}{\medskip}

\newenvironment{theo}{\refstepcounter{definition} \medskip
\noindent{\it Theorem \thedefinition:}}{\medskip}
\newenvironment{prop}{\refstepcounter{definition} \medskip
\noindent{\it Proposition \thedefinition:}}{\medskip}
\newenvironment{lem}{\refstepcounter{definition} \medskip
\noindent{\it Lemma \thedefinition:}}{\medskip}

\newcommand{\mar}[1]{}

\begin{document}
\hbox{}

{\parindent=0pt

{\Large \bf Geometric quantization of mechanical systems with
time-dependent parameters}

\bigskip

{\sc G.Giachetta}\footnote{Electronic mail: giachetta@campus.unicam.it},
{\sc L.Mangiarotti}\footnote{Electronic mail: mangiaro@camserv.unicam.it}

{\sl Department of Mathematics and Physics, University of Camerino,
62032 Camerino (MC), Italy}

\medskip

{\sc G. Sardanashvily}\footnote{Electronic mail:
sard@grav.phys.msu.su}

{\sl Department of Theoretical Physics,
Moscow State University, 117234 Moscow, Russia}

\bigskip

Quantum systems with
adiabatic classical parameters are widely studied, e.g., in the
modern holonomic quantum computation. 
We here provide complete geometric quantization
of a Hamiltonian system with time-dependent parameters, without the adiabatic
assumption. A Hamiltonian of such a system is
affine in the temporal derivative of parameter functions. This
leads to the geometric Berry factor phenomena.

}

\bigskip

\noindent
{\bf I. INTRODUCTION}
\bigskip

At present, quantum systems with classical parameters attract special 
attention in connection with holonomic quantum computation.$^{1-3}$
This approach to quantum computing is based on the generalization 
of Berry's adiabatic phase to the nonabelian case corresponding to 
adiabatically driving an $n$-fold degenerate eigen-state of a Hamiltonian
over the
parameter manifold.$^4$ In the framework of 
the holonomic quantum computation scheme, information is encoded in
this degenerate state, while 
the parameter manifold plays the role of a control parameter space.
The key point of this scheme is that the parallel transport along the
curves in the parameter space is assumed
to be adiabatic
with respect to a dynamic Hamiltonian. 

At the same time, the adiabatic condition of  Berry's phase phenomena
can be removed.$^{5,6}$ Moreover, one observes that the
Berry factor is a standard ingredient in evolution of 
quantum systems with classical time-dependent parameters.$^{7,8}$ Here,
we provide complete geometric quantization of a mechanical
system depending on parameters, without adiabatic assumption.

The configuration space of such a system is a composite fiber bundle
\mar{pr1}\beq
Q\to \Si \to \bR, \label{pr1}
\eeq
where $\bR$ is the time axis and $\Si\to\bR$ is a fiber bundle
whose sections are parameter functions.$^{7-9}$ The configuration space
(\ref{pr1}) is coordinated by
$(t,\si^\la,q^k)$, where $(t,\si^\la)$ are bundle coordinates on $\Si\to\bR$
and $t$ is a fixed Cartesian coordinate on $\bR$.
The corresponding momentum phase space is the vertical cotangent bundle
$V^*Q$ of $Q\to\Si$ equipped with holonomic coordinates
$(t,\si^\la,q^k,p_k)$. It is provided with the
following canonical Poisson structure.
Let $T^*Q$ be the cotangent
bundle of $Q$ endowed with the canonical symplectic form $\Om_T$ and the
associated Poisson bracket $\{,\}_T$. Given the canonical fibration
\mar{lmp44}\beq
\zeta: T^*Q\to V^*Q, \label{lmp44}
\eeq
the Poisson bracket $\{,\}_V$ on the $\bR$-ring
$C^\infty(V^*Q)$ of smooth real functions on $V^*Q$ is defined by the relation
\mar{m72',2}\ben
&& \zeta^*\{f,f'\}_V=\{\zeta^*f,\zeta^*f'\}_T, \qquad  \label{m72'} \\
&& \{f,f'\}_V = \dr^kf\dr_kf'-\dr_kf\dr^kf', \qquad
f,f'\in C^\infty(V^*Q). \label{m72}
\een
Its characteristic symplectic foliation coincides with the fibration
\mar{pr2}\beq
\pi_{V\Si}:V^*Q\to\Si. \label{pr2}
\eeq
One can think of its fiber $V^*_\si Q=T^*Q_\si$, $\si\in \Si$, as being the
momentum phase space of a mechanical system under fixed values of parameters
and at a given instant of time.

It seems natural to quantize the Poisson manifold $(V^*Q,\{,\}_V)$ in 
accordance
with the well-known geometric
quantization procedure,$^{10,11}$ but then one faces the problem that
the mean values of quantum operators are defined via integration over
the whole momentum phase space $V^*Q$, including integration
over classical parameters.
At the same time, quantization
of a system with classical parameters should necessarily imply its quantization
under fixed values of parameters. Its generic carrier space
is a Hilbert module of
sections of a smooth Hilbert bundle over the parameter
space $\Si$,$^{7,8}$ i.e., a locally-trivial smooth field of Hilbert
spaces on $\Si$ in the terminology of Ref. [12]. In particular, the
instantwise quantization of nonrelativistic time-dependent mechanics
(where $\Si=\bR$) is of this type.$^{13,14}$
However, geometric quantization of a Poisson manifold need not yield
quantization of its symplectic leaves.$^{15}$

To dispose of these problems, we will apply to $V^*Q\to\Si$ the
technique of leafwise
geometric quantization of symplectic foliations.$^{16}$ There is one-to-one
correspondence between Poisson structures on a smooth manifold and its
symplectic foliations.
The quantum algebra of a symplectic
foliation is a particular quantum algebra of the associated Poisson
manifold such that its restriction to each symplectic leaf is defined
and quantized.
We choose the canonical real polarization of $V^*Q\to \Si$ which is
the vertical tangent bundle
$VV^*Q$ of the fiber bundle $V^*Q\to Q$. The corresponding
quantum algebra $\cA_\cF$ consists of functions on $V^*Q$ which are affine
in momenta $p_k$. It is represented by Schr\"odinger operators in the
pre-Hilbert $C^\infty(\Si)$ module $E_Q$ of fiberwise complex
half-forms on the fiber
bundle $Q\to\Si$ whose restriction to each fiber is of compact support.
This representation is naturally extended to the enveloping algebra
$\ol\cA_\cF$ of polynomial functions in momenta.

We show that a Hamiltonian of a quantum system with classical parameters is
affine in the temporal derivative $\dr_t\chi^\la$ of parameter functions
$\chi^\la$, namely,
\be
\wh\cH=-i(\La^k_\la\dr_k +\frac12 \dr_k\La^k_\la)\dr_t\chi^\la + 
\wh \cH'(\chi),
\ee
where $\La^k_\la$ are components of a connection on the fiber bundle 
$Q\to\Si$. The key point is that integration of the first term 
of this Hamiltonian
over time through a parameter function $\chi(t)$ depends only on a
trajectory of this function in a parameters space, but not on
parametrization of this trajectory by time (i.e., the adiabatic
assumption is not necessary). As a consequence, this term 
is responsible for the geometric Berry factor phenomena. 
It plays the role of a control operator in holonomic
quantum computation.

\bigskip

\noindent
{\bf II. THE LEAFWISE DIFFERENTIAL CALCULUS}
\bigskip

Geometric quantization of symplectic foliations is phrased in terms of
the leafwise differential calculus on a
foliated manifold.
Manifolds throughout are assumed to be smooth, Hausdorff,
second-countable (i.e., paracompact), and connected.

Recall that a (regular) foliation $\cF$ of a manifold $Z$ consists
of (maximal) integral manifolds of an involutive distribution
$i_\cF:T\cF\to TZ$ on $Z$. A foliated manifold
$(Z,\cF)$ admits an adapted coordinate atlas
\mar{spr850}\beq
\{(U_\xi;z^\la; z^i)\},\quad \la=1,\ldots,{\rm codim}\,\cF, \quad
i=1,\ldots,\di\cF,
\label{spr850}
\eeq
such that transition functions of coordinates $z^\la$ are independent of the
remaining coordinates $z^i$ and,
for each leaf $F\in\cF$, the connected components
of $F\cap U_\xi$ are given by the equations
$z^\la=$const. These connected components and coordinates $(z^i)$ on
them make up a coordinate atlas of a leaf $F$.

The real Lie algebra $\cT_1(\cF)$ of
global sections of the distribution $T\cF\to Z$ is
a $C^\infty(Z)$-submodule of the
derivation module of the $\bR$-ring
$C^\infty(Z)$ of smooth real functions on $Z$. Its kernel
$S_\cF(Z)\subset C^\infty(Z)$ consists of functions constant
on leaves of $\cF$. Therefore,
$\cT_1(\cF)$ is the Lie $S_\cF(Z)$-algebra of derivations of $C^\infty(Z)$,
regarded as a $S_\cF(Z)$-ring.
Then one can introduce the leafwise differential
calculus$^{17}$ as the
Chevalley--Eilenberg differential calculus
over the $S_\cF(Z)$-ring $C^\infty(Z)$. It is a subcomplex
\mar{spr892}\beq
0\to S_\cF(Z)\ar C^\infty(Z)\ar^{\wt d} \gF^1(Z) \cdots
\ar^{\wt d} \gF^{\di\cF}(Z) \to 0 \label{spr892}
\eeq
of the Chevalley--Eilenberg complex of
the Lie $S_\cF(Z)$-algebra $\cT_1(\cF)$
with coefficients in $C^\infty(Z)$ which consists of
$C^\infty(Z)$-multilinear skew-symmetric maps
$\op\times^r \cT_1(\cF) \to C^\infty(Z)$, $r=1,\ldots,\di\cF$.$^8$
These maps are global sections of exterior products
$\op\w^r T\cF^*$ of the dual $T\cF^*\to Z$ of $T\cF\to Z$.
They are called the leafwise forms
on a foliated manifold $(Z,\cF)$, and are given by the coordinate expression
\be
\f=\frac1{r!}\f_{i_1\ldots i_r}\wt dz^{i_1}\w\cdots\w \wt dz^{i_r},
\ee
where $\{\wt dz^i\}$ are the duals of the holonomic fiber
bases $\{\dr_i\}$ for $T\cF$.
Then one can think of the
Chevalley--Eilenberg coboundary operator
\be
\wt d\f= \wt dz^k\w \dr_k\f=\frac{1}{r!}
\dr_j\f_{i_1\ldots i_r}\wt dz^j\w\wt dz^{i_1}\w\cdots\w\wt dz^{i_r}
\ee
as being the leafwise exterior differential.
Accordingly, the complex $(S_\cF(Z),\gF^*(Z),\wt d)$ (\ref{spr892}) 
is called the
leafwise de Rham complex
(or the tangential de Rham complex
in the terminology of Ref. [17]). This is the complex
$(\cA^{0,*},d_f)$ in Ref. [18]. Its cohomology $H^*_\cF(Z)$ is called
the leafwise de Rham cohomology.

Let us consider the exact sequence
\mar{spr917}\beq
  0\to {\rm Ann}\,T\cF\ar T^*Z\ar^{i^*_\cF} T\cF^* \to 0 \label{spr917}
\eeq
of vector bundles over $Z$. Since it is always split,
the epimorphism $i^*_\cF$ yields an
epimorphism  of the graded algebra $\cO^*(Z)$ of exterior forms on $Z$ to
the algebra $\gF^*(Z)$ of leafwise forms. The relation
$i^*_\cF\circ d=\wt d\circ i^*_\cF$ holds and, thereby,
we have the cochain morphism
\mar{lmp04}\beq
  i^*_\cF: (\bR,\cO^*(Z),d)\to (S_\cF(Z),\gF^*(Z),\wt d), \qquad
  dz^\la\mapsto 0, \quad dz^i\mapsto\wt dz^i, \label{lmp04}
\eeq
of the de Rham complex of $Z$
to the leafwise de Rham complex (\ref{spr892}), and the
corresponding homomorphism
\mar{lmp00}\beq
[i^*_\cF]^*: H^*(Z)\to H^*_\cF(Z) \label{lmp00}
\eeq
of the de Rham cohomology of $Z$ to the leafwise one.

Given a leaf $i_F:F\to Z$ of a foliation $\cF$, there is the pull-back
homomorphism
\mar{lmp30}\beq
(\bR,\cO^*(Z),d) \to (\bR,\cO^*(F),d) \label{lmp30}
\eeq
of the de Rham complex of $Z$ to that of $F$ and the corresponding
homomorphism of their cohomology groups
\mar{lmp31}\beq
H^*(Z) \to H^*(F). \label{lmp31}
\eeq

\begin{prop} \label{lmp32} \mar{lmp32} The homomorphisms (\ref{lmp30}) and
(\ref{lmp31}) factorize through the homomorphisms (\ref{lmp04}) and
(\ref{lmp00}), respectively.
\end{prop}

\begin{proof}
It is readily observed that
the pull-back bundles $i_F^* T\cF$ and $i_F^* T\cF^*$ over $F$ are
isomorphic to the tangent and the cotangent
bundles of $F$, respectively. Moreover, a direct computation shows that
$i_F^*(\wt d\f)=d(i_F^*\f)$ for
any leafwise form $\f$. It follows that the cochain morphism (\ref{lmp30})
factorizes through the cochain morphism
(\ref{lmp04}) and the cochain morphism
\be
i^*_F: (S_\cF(Z),\gF^*(Z),\wt d)\to (\bR,\cO^*(F),d), \qquad
\wt dz^i\mapsto dz^i,
\ee
of the leafwise de Rham complex of $(Z,\cF)$ to the de Rham complex
of $F$. Accordingly, the cohomology morphism (\ref{lmp31})
factorizes through the leafwise cohomology
\be
H^*(Z)\ar^{[i^*_\cF]} H^*_\cF(Z)\ar^{[i_F^*]} H^*(F).
\ee
\end{proof}

Turn now to symplectic foliations.
Let $\cF$ be of even dimension.
A $\wt d$-closed non-degenerate leafwise
two-form $\Om_\cF$ on a foliated manifold $(Z,\cF)$ is called
symplectic. Its pull-back $i_F^*\Om_\cF$ onto
each leaf $F$ of $\cF$ is a
symplectic form on $F$. A leafwise symplectic form $\Om_\cF$ yields the
bundle isomorphism
\be
\Om_\cF^\fl: T\cF\op\to_Z T\cF^*, \qquad
\Om_\cF^\fl:v\mapsto - v\rfloor\Om_\cF(z),
\qquad v\in T_z\cF.
\ee
The inverse isomorphism $\Om_\cF^\sh$ determines the
Poisson bivector field
\mar{spr904}\beq
w_\Om(\al,\bt)=\Om_\cF(\Om_\cF^\sh(i^*_\cF\al),\Om_\cF^\sh(i^*_\cF\bt)),
\qquad \forall \al,\bt\in T_z^*Z, \quad z\in Z,
\label{spr904}
\eeq
on $Z$ subordinate to $\op\w^2T\cF$.
The corresponding Poisson bracket reads
\mar{spr902}\beq
\{f,f'\}_\cF=\vt_f\rfloor \wt df', \qquad
\vt_f\rfloor\Om_\cF=-\wt df, \qquad \vt_f=\Om_\cF^\sh(\wt df),
\qquad f,f'\in C^\infty(Z).\label{spr902}
\eeq
Its kernel is $S_\cF(Z)$.
Conversely, let $(Z,w)$ be a (regular) Poisson manifold and $\cF$ its
characteristic foliation. Since Ann$\,T\cF\subset T^*Z$ is
precisely the kernel of the Poisson bivector field $w$,
the bundle homomorphism $w^\sh: T^*Z\op\to_Z TZ$
factorizes in a unique fashion
\be
w^\sh:
T^*Z\ar_Z^{i^*_\cF} T\cF^*\ar_Z^{w^\sh_\cF}
T\cF\ar_Z^{i_\cF} TZ
\ee
through the bundle isomorphism
\be
w_\cF^\sh: T\cF^*\op\to_Z T\cF,  \qquad
w^\sh_\cF:\al\mapsto -w(z)\lfloor \al, \qquad
\al\in T_z\cF^*.
\ee
The inverse isomorphism $w_\cF^\fl$ yields the symplectic leafwise form
\mar{spr903}\beq
\Om_\cF(v,v')=w(w_\cF^\fl(v),w_\cF^\fl(v')), \qquad \forall v,v'\in T_z\cF,
\qquad z\in Z. \label{spr903}
\eeq
The formulae (\ref{spr904}) and (\ref{spr903}) establish the
above-mentioned equivalence
between the Poisson structures on a manifold $Z$
and its symplectic foliations, but this equivalence need not be
preserved under morphisms.

\bigskip

\noindent
{\bf III. PREQUANTIZATION OF SYMPLECTIC FOLIATIONS}
\bigskip

Prequantization of a symplectic foliation $(\cF,\Om_\cF)$ of a manifold
$Z$ provides a representation
\mar{lqm514}\beq
f\mapsto i\wh f, \qquad [\wh f,\wh f']=-i\wh{\{f,f'\}}_\cF, \label{lqm514}
\eeq
of the Poisson algebra $(C^\infty(Z),\{f,f'\}_\cF)$ by first order
differential operators on
sections of a complex line
bundle $\pi:C\to Z$. These operators
are given by the
Kostant--Souriau formula
\mar{lqq46}\beq
\wh f=-i\nabla_{\vt_f}^\cF +\ve f,
\qquad \ve>0, \label{lqq46}
\eeq
where $\vt_f$ is the Hamiltonian vector field (\ref{spr902}) and
$\nabla^\cF$ is the covariant differential with respect to a
leafwise connection on $C\to Z$ whose
curvature form obeys the prequantization
condition
\mar{lmp61}\beq
\wt R=i\ve\Om_\cF. \label{lmp61}
\eeq
  In this Section, we
provide the cohomology analysis of this condition, and show
that prequantization of a symplectic foliation
yields prequantization of its symplectic leaves.
The key point is that any leafwise connection comes from a connection
(see Theorem \ref{lmp42} below).

The inverse
images $\pi^{-1}(F)$ of leaves $F$ of the foliation $\cF$ of $Z$
make up a (regular) foliation $C_\cF$ of a complex line bundle $C$.
Given the (holomorphic)
tangent bundle $TC_\cF$ of this foliation, we have the
exact sequence of vector bundles
\mar{lmp18}\beq
0\to VC\ar_C TC_\cF\ar_C C\op\times_Z T\cF\to 0, \label{lmp18}
\eeq
where $VC$ is the (holomorphic) vertical tangent bundle of $C\to Z$.
A (linear) leafwise
connection on the complex line bundle $C\to Z$ is defined as a splitting
of the exact sequence (\ref{lmp18}) which is linear over $C$.

One can choose an adapted coordinate atlas $\{(U_\xi;z^\la; z^i)\}$
(\ref{spr850}) of a foliated manifold $(Z,\cF)$ such that $U_\xi$ are
trivialization domains of the complex line bundle $C\to Z$. Let
$(z^\la; z^i;c)$, $c\in\bC$,
be the
corresponding bundle coordinates on $C\to Z$.
They are also adapted coordinates on the
foliated manifold $(C,C_\cF)$.
With respect to these coordinates, a leafwise connection
is represented by a $TC_\cF$-valued leafwise
one-form
\mar{lmp21}\beq
A_\cF=\wt dz^i\ot(\dr_i +A_ic\dr_c), \label{lmp21}
\eeq
where $A_i$ are local complex functions on $C$.
The covariant differential on the $C^\infty(Z)$-module
$C(Z)$ of sections the line bundle
$C\to Z$ with respect to the leafwise connection (\ref{lmp21}) reads
\be
\nabla^\cF s=\wt ds- A_i s\wt dz^i. \qquad s\in C(Z).
\ee

The exact sequence (\ref{lmp18}) is obviously a subsequence of the
exact sequence
\be
0\to VC\ar_C TC\ar_C C\op\times_Z TZ\to 0,
\ee
where $TC$ is the holomorphic tangent bundle of $C$.
Consequently, any connection
\mar{lmp103}\beq
\G=dz^\la\ot(\dr_\la + \G_\la c\dr_c) +
dz^i\ot(\dr_i +\G_ic\dr_c) \label{lmp103}
\eeq
on a complex line bundle $C\to Z$ yields a leafwise connection
\mar{lmp23}\beq
\G_\cF=\wt dz^i\ot(\dr_i +\G_ic\dr_c). \label{lmp23}
\eeq

\begin{theo} \label{lmp42} \mar{lmp42}
Any leafwise connection on a complex line bundle $C\to Z$ comes from
a connection on it.
\end{theo}

\begin{proof}
Let $A_\cF$ (\ref{lmp21}) be a leafwise connection on
$C\to Z$ and $\G_\cF$
(\ref{lmp23}) a leafwise connection which comes from some connection
$\G$ (\ref{lmp103}) on
$C\to Z$. Their affine difference over $C$ is a section
\be
Q=A_\cF-\G_\cF=\wt dz^i\ot(A_i -\G_i)c\dr_c
\ee
of the vector bundle $T\cF^*\op\ot_CVC\to C$.
Given some splitting
\be
B: \wt dz^i \mapsto dz^i- B^i_\la dz^\la
\ee
of the exact sequence (\ref{spr917}), the composition
\be
(B\ot \id_{VC})\circ Q=(dz^i- B^i_\la dz^\la)\ot(A_i -\G_i)c\dr_c: C\to
T^*Z\op\ot_C VC
\ee
is a soldering form on the complex line bundle $C\to Z$. Then
\be
\G+(B\ot\id_{VC})\circ Q=
dz^\la\ot(\dr_\la + [\G_\la -B^i_\la (A_i -\G_i)]c\dr_c) +
dz^i\ot(\dr_i +A_ic\dr_c)
\ee
is a desired connection on $C\to Z$ which yields the leafwise
connection $A_\cF$ (\ref{lmp21}).
\end{proof}

The curvature of the leafwise connection $A_\cF$ (\ref{lmp21}) is defined as a
$C^\infty(Z)$-linear endomorphism
\be
\wt R(\tau,\tau')=\nabla_{[\tau,\tau']}^\cF- [\nabla_\tau^\cF,
\nabla_{\tau'}^\cF]=\tau^i
\tau'^j R_{ij}, \qquad
R_{ij}=\dr_i A_j-\dr_j A_i,
\ee
of $C(Z)$ for any vector fields $\tau,\tau'\in \cT_1(\cF)$.
It is represented by the
complex leafwise two-form
\mar{lmp08}\beq
\wt R=\frac12 R_{ij}\wt dz^i\w \wt dz^j. \label{lmp08}\\
\eeq
If a leafwise connection $A_\cF$ comes from a connection $A$,
its curvature form $\wt R$ (\ref{lmp08}) is the image
$\wt R=i^*_\cF R$
of the curvature form $R$ of $A$ with respect to the
morphism $i^*_\cF$ (\ref{lmp04}).

Let us return to the prequantization condition (\ref{lmp61}).

\begin{lem} \label{lmp63} \mar{lmp63}
Let us assume that there is a leafwise connection
$\G_\cF$ on a complex line bundle $C\to Z$ which fulfils the
prequantization condition (\ref{lmp61}). Then,
for any Hermitian metric $g$ on $C\to Z$, there exists
a leafwise connection $A_\cF^g$ on $C\to Z$ which: (i) satisfies the
condition
(\ref{lmp61}), (ii) preserves $g$, and (iii) comes from a
$U(1)$-principal connection on $C\to Z$.
\end{lem}

\begin{proof}
For any Hermitian metric $g$ on $C\to Z$, there
exists an associated bundle atlas $\Psi^g=\{(z^\la;z^i,c)\}$ of $C$ with
$U(1)$-valued transition functions such that 
\mar{lll}\beq
g(c,c')=c\ol c'. \label{lll}
\eeq
Let the above mentioned leafwise connection $\G_\cF$ come from
a linear connection $\G$ (\ref{lmp103})
on $C\to Z$ written with respect to the atlas $\Psi^g$.
The connection $\G$ is split into the sum $A^g + \g$ where,
\mar{lmp62}\beq
A^g=dz^\la\ot(\dr_\la + {\rm Im}(\G_\la)c\dr_c) +
dz^i\ot(\dr_i +{\rm Im}(\G_i) c\dr_c) \label{lmp62}
\eeq
is a $U(1)$-principal connection, preserving the Hermitian metric $g$. The
curvature forms $R$ of $\G$ and $R^g$ of $A^g$
obey the relation $R^g={\rm Im}(R)$.
The connection $A^g$ (\ref{lmp62}) defines the leafwise
connection
\mar{lmp73}\beq
A_\cF^g=i_\cF^*A^g= \wt dz^i\ot(\dr_i + iA^g_i c\dr_c), \qquad iA^g_i=
{\rm Im}(\G_i), \label{lmp73}
\eeq
preserving the Hermitian metric $g$ (\ref{lll}).
Its curvature fulfils the desired
relation
\mar{lmp65}\beq
\wt R^g=i_\cF^*R^g={\rm Im}(i_\cF^*R)= i\ve\Om_\cF. \label{lmp65}
\eeq
\end{proof}

Since $A^g$ (\ref{lmp62}) is a $U(1)$-principal connection, its
curvature form $R^g$ is related to the first Chern form
of integer de Rham cohomology class
by the formula
$c_1=i(2\pi)^{-1}R^g$.
If the prequantization
condition (\ref{lmp61}) holds,
the relation (\ref{lmp65}) shows that the leafwise
cohomology class of the leafwise
form $(2\pi)^{-1}\ve\Om_\cF$ is the image of an integer de Rham
cohomology class with respect to the cohomology morphism $[i^*_\cF]$
(\ref{lmp00}). Conversely, if a leafwise
symplectic form $\Om_\cF$ on a foliated manifold $(Z,\cF)$ is of this
type, there exists a complex
line bundle $C\to Z$ and a $U(1)$-principal connection $A$ on $C\to Z$
such that the leafwise connection $i^*_\cF A$ fulfils the relation
(\ref{lmp61}). Thus, we have stated the following.

\begin{prop} \label{lmp66} \mar{lmp66}
A symplectic foliation $(\cF,\Om_\cF)$ of a manifold $Z$ admits
prequantization (\ref{lqq46}) iff
the leafwise cohomology class of $(2\pi)^{-1}\ve\Om_\cF$ is the image
of an integer de Rham
cohomology class of $Z$.
\end{prop}

The leafwise connection $A^g_\cF$ in Lemma \ref{lmp63}
by no means is unique. Let
$\f=\f_i\wt dz^i$ be a closed leafwise one-form. Then the connection
\be
A'^g_\cF=\wt dz^i\ot(\dr_i + i(A^g_i+\f_i) c\dr_c)
\ee
also obeys the prequantization condition (\ref{lmp08}) and preserves the
Hermitian metric (\ref{lll}).

Let $F$ be a leaf of a symplectic foliation
$(\cF,\Om_\cF)$ provided with the symplectic form $\Om_F=i^*_F\Om_\cF$.
In accordance with Proposition \ref{lmp32},
the symplectic form $(2\pi)^{-1}\ve \Om_F$ belongs to an integer
de Rham cohomology
class if a leafwise symplectic form $\Om_\cF$ fulfils the condition of
Proposition \ref{lmp66}. Thus, if a symplectic foliation
admits prequantization, its symplectic leaves
do as well.
The corresponding prequantization bundle for $F$ is the pull-back
complex line bundle $i^*_FC$, coordinated by $(z^i,c)$. Furthermore,
let $A_\cF^g$ (\ref{lmp73}) be
a leafwise connection on the prequantization bundle $C\to Z$
which obeys Lemma \ref{lmp63}, i.e., comes from a
$U(1)$-principal connection $A^g$ on $C\to Z$. Then the pull-back
\mar{lmp130}\beq
A_F=i^*_FA^g=dz^i\ot(\dr_i +ii^*_F(A^g_i)c\dr_c) \label{lmp130}
\eeq
of the connection $A^g$ onto
$i^*_FC\to F$ satisfies the
prequantization condition
\be
R_F=i^*_FR=i\ve \Om_F,
\ee
and preserves the pull-back Hermitian metric $i^*_Fg$ on $i^*_\cF C\to
F$.

\bigskip

\noindent
{\bf IV. QUANTIZATION OF SYMPLECTIC FOLIATIONS}
\bigskip

The next step is polarization of a symplectic foliation $(\cF,\Om_\cF)$ of
a manifold $Z$. It is defined as a maximal involutive distribution
$\bT\subset T\cF$ on $Z$ such that
\be
\Om_\cF(u,v)=0, \qquad \forall u,v\in\bT_z, \qquad z\in Z.
\ee
Given  the Lie algebra $\bT(Z)$ of $\bT$-subordinate vector fields on
$Z$, the quantum algebra of $(\cF,\Om_\cF)$ is defined as the complexified
subalgebra $\cA_\cF\subset C^\infty(Z)$ of functions $f$
whose Hamiltonian vector fields $\vt_f$ (\ref{spr902}) fulfil
the condition $[\vt_f,\bT(Z)]\subset \bT(Z)$.
This algebra obviously contains the center $S_\cF(Z)$
of the Poisson algebra $(C^\infty(Z),\{,\}_\cF)$, and is a Lie
$S_\cF(Z)$-algebra.

Let $(F,\Om_F)$ be a symplectic leaf of a symplectic foliation
$(\cF,\Om_\cF)$. Given a polarization $\bT\to Z$ of $(\cF,\Om_\cF)$, its
restriction $\bT_F=i^*_F\bT\subset i^*_FT\cF=TF$ to $F$ is an
involutive distribution on $F$. It obeys the condition
\be
i^*_F\Om_\cF(u,v)=0, \qquad \forall u,v\in\bT_{Fz}, \qquad z\in F,
\ee
i.e., it is a polarization of the symplectic manifold $(F,\Om_F)$.
Thus, polarization of a symplectic foliation defines polarization of each
symplectic leaf.
Clearly, the quantum algebra $\cA_F$ of a symplectic leaf $F$ with respect to
the polarization $\bT_F$ contains all functions $i^*_Ff$ of the quantum algebra
$\cA_\cF$ restricted to $F$.

Let us note that every polarization $\bT$ of a symplectic foliation 
$(\cF,\Om_\cF)$
yields polarization of the associated Poisson manifold $(Z,\{,\}_\cF)$.
It is the sheaf  $\Phi$ of germs of local functions $f$
on $Z$ whose Hamiltonian
vector fields $\vt_f$ (\ref{spr902}) are subordinate to $\bT$, i.e.
their Poisson bracket vanishes.
Since the cochain morphism $i^*_\cF$ (\ref{lmp04}) is
an epimorphism, the leafwise differential calculus $\gF^*(Z)$ is
universal, i.e., the leafwise differentials $\wt df$ of functions $f\in
C^\infty(Z)$ on
$Z$ make up a basis for the $C^\infty(Z)$-module $\gF^1(Z)$. Let
$\Phi(Z)$ denote the structure $\bR$-module of global sections of the
sheaf $\Phi$. Then the leafwise differentials of elements of $\Phi(Z)$
make up a basis for the $C^\infty(Z)$-module of global sections of the
codistribution $\Om_\cF^\fl\bT$. Equivalently, the Hamiltonian
vector fields of elements of $\Phi(Z)$ constitute a basis for the
$C^\infty(Z)$-module $\bT(Z)$. One can easily show that
polarization $\bT$ of a symplectic foliation $(\cF,\Om_\cF)$ and the
corresponding polarization $\Phi$ of the Poisson manifold $(Z,w_\Om)$
define the same quantum algebra $\cA_\cF$.

Though $\cA_\cF$ coincides with the quantum algebra of the Poisson
manifold $(Z,\{,\}_\cF)$, we modify the
standard metaplectic correction technique$^{19,20}$ as follows in order to
provide the leafwise quantization of $\cA_\cF$.$^{16}$

Let us consider the exterior bundle $\op\w^mT\cF^*$, $m=\di\cF$.
Its structure group $GL(m,\bR)$
is reducible to the group $GL^+(m,\bR)$ since a symplectic
foliation is oriented. One can regard this fiber bundle as being
associated to a $GL(m,\bC)$-principal bundle $P\to Z$.
Let us assume that $H^2(Z;\bZ_2)=0$. Then the principal
bundle $P$ admits a two-fold covering principal bundle with the
structure metalinear group $ML(m,\bC)$.$^{20}$
As a consequence, there
exists a complex line bundle $\cD_\cF\to Z$
characterized by an atlas $\Psi_\cF=\{(U_\xi;z^\la;z^i;c)\}$ with
the transition functions $c'=S_\cF c$ such that
\be
S_\cF\ol S_\cF=\det\left(\frac{\dr z^i}{\dr z'^j}\right).
\ee
One can think of its sections as being leafwise half-forms on $Z$.
The metalinear bundle $\cD_\cF\to Z$ admits the canonical lift
of any $\bT$-subordinate vector field $\tau$ on $Z$.
The corresponding
Lie derivative of its sections reads
\mar{lmp82}\beq
\bL_\tau=\tau^i\dr_i+\frac12\dr_i\tau^i. \label{lmp82}
\eeq

We define the quantization bundle as the tensor product
$Y=C\ot\cD_\cF$.
Given a leafwise connection $A^g_\cF$ (\ref{lmp73})
and the Lie derivative $\bL$ (\ref{lmp82}), let us
associate  the first order differential operator
\mar{lmp84}\beq
\wh f=-i[(\nabla_{\vt_f}^\cF +i\ve f)\ot\id +\id\ot\bL_{\vt_f}]=
-i[\nabla_{\vt_f}^\cF +i\ve f+\frac12\dr_i\vt_f^i], \qquad f\in\cA_\cF,
\label{lmp84}
\eeq
on sections $\rho$ of $Y$ to
each element of the quantum algebra $\cA_\cF$. A direct
computation with respect to the local Darboux coordinates on $Z$ shows
that the operators (\ref{lmp84}) obey the Dirac condition (\ref{lqm514})
and that, if a section $\rho$ fulfils the relation
\mar{lmp86}\beq
(\nabla_\vt^\cF\ot\id +\id\ot\bL_\vt)\rho=(\nabla_\vt^\cF
+\frac12\dr_i\vt^i)\rho=0  \label{lmp86}
\eeq
for all $\bT$-subordinate Hamiltonian vector field $\vt$,
then $\wh f\rho$ possesses the same property for any $f\in\cA_\cF$.

Let us restrict the representation of the
quantum algebra $\cA_\cF$ by the operators (\ref{lmp84}) to the subspace
$E\in Y(Z)$ of
sections $\rho$ which obey the condition (\ref{lmp86}) and whose
restriction to
any leaf of $\cF$ is of compact support. The last condition is
motivated by the following.

Since $i^*_FT\cF^*=T^*F$, the pull-back $i^*_F\cD_\cF$
of $\cD_\cF$ onto a leaf $F$ is a metalinear
bundle of half-forms on $F$. Therefore, the pull-back
$i^*_FY$ of the quantization bundle $Y\to Z$ onto
$F$ is a quantization bundle for the symplectic manifold
$(F,i^*_F\Om_\cF)$. Given the pull-back connection $A_F$ (\ref{lmp130})
and the polarization $\bT_F=i^*_F\bT$, this symplectic manifold is subject to
the standard geometric quantization by the
first order differential operators
\mar{lmp92}\beq
\wh f=-i(i_F^*\nabla_{\vt_f}^\cF +i\ve f +\frac12\dr_i\vt_f^i), \qquad
f\in \cA_F,  \label{lmp92}
\eeq
on sections $\rho_F$ of $i^*_FY\to F$ of compact support
which obey the condition
\be
(i_F^*\nabla_\vt^\cF +\frac12\dr_i\vt^i)\rho_F=0
\ee
for all $\bT_F$-subordinate Hamiltonian vector fields $\vt$ on $F$.
These sections constitute a pre-Hilbert space $E_F$ with respect to
the Hermitian form
\be
\lng\rho_F|\rho'_F\rng=\left(\frac1{2\pi}\right)^{m/2}\op\int_F
\rho_F \ol\rho'_F,
\ee
and $\wh f$ (\ref{lmp92}) are Hermitian operators in $E_F$.
It is readily observed that
$i^*_FE\subset E_F$ and, moreover, the relation
\be
i^*_F(\wh f\rho)=\wh{(i^*_Ff)}(i^*_F\rho)
\ee
holds for all elements $f\in\cA_\cF$ and $\rho\in E$.
This relation enables one to think of the operators $\wh f$
(\ref{lmp84}) in $E$ as being the leafwise quantization of the
$S_\cF(Z)$-algebra $\cA_\cF$ by Hermitian operators
in a pre-Hilbert $S_\cF(Z)$-module.

\bigskip

\noindent
{\bf V. QUANTIZATION OF A MECHANICAL SYSTEM WITH PARAMETERS}
\bigskip

Let $Q$ (\ref{pr1}) be the configuration space of a mechanical system
with parameters. We assume that $H^2(Q;\bZ_2)= H^2(V^*Q;\bZ_2)=0$.
The characteristic symplectic foliation $\cF$ of the Poisson structure
(\ref{m72}) on the momentum phase space $V^*Q$
is the fibration $\pi_{V\Si}$ (\ref{pr2}) endowed
with the leafwise symplectic form
\be
\Om_\cF=\wt dp_k\w\wt dq^k.
\ee
Since this form is $\wt d$-exact, its leafwise de Rham cohomology class
equals zero and is the image of the zero de Rham cohomology class with
respect to the morphism $[i^*_\cF]$ (\ref{lmp00}).
Then, in accordance with Proposition
\ref{lmp66}, the symplectic foliation $(V^*Q\to\Si,\Om_\cF)$ admits
prequantization.

The prequantization bundle $C\to V^*Q$, associated to
the zero Chern class, is trivial. Let its trivialization $C=V^*Q\times
\bC$ hold fixed, and let $(t,\si^\la,q^k,p_k,c)$ be the corresponding
bundle coordinates. Unless $Q$ is specified, we choose the leafwise connection
\be
A_\cF=\wt dp_k\ot\dr^k + \wt dq^k\ot(\dr_k+ip_kc\dr_c)
\ee
on $C\to V^*Q$. This connection preserves the Hermitian metric
$g$ (\ref{lll}) on $C$, and its curvature fulfils the 
prequantization condition
$\wt R=i\Om_\cF$. The corresponding prequantization operators
(\ref{lqq46}) read
\be
\wh f=-i\vt_f +(f-p_k\dr^kf), \qquad \vt_f=\dr^kf\dr_k-\dr_kf\dr^k,
\qquad f\in C^\infty(V^*Q).
\ee

Let us choose the canonical vertical polarization of the symplectic foliation
$(V^*Q\to\Si,\Om_\cF)$ which is the vertical tangent bundle
$\bT=VV^*Q$ of the
fiber bundle
\be
\pi_{VQ}:V^*Q\to Q.
\ee
It is readily observed that the corresponding quantum algebra
$\cA_\cF\subset C^\infty(V^*Q)$
consists of functions which are affine in momenta coordinates $p_k$.
Due to the linear transformation law of $p_k$, this property is
coordinate-independent.

Following the quantization procedure in the previous Section, one should
define a representation of $\cA_\cF$ in the space $E$ of sections
$\rho$ of
the quantization bundle $C\ot\cD_\cF$ which obey the condition
(\ref{lmp86}) and whose restriction to each fiber of
$V^*Q\to \Si$ is of
compact support. The condition (\ref{lmp86}) reads
\be
\dr_kf\dr^k\rho=0, \qquad \forall f\in C^\infty(Q),
\ee
i.e., elements of $E$ are constant on fibers of $V^*Q\to Q$. Consequently,
$E$ is $\{0\}$.

Therefore, we modify the quantization procedure as follows. Given
the exterior bundle $\op\w^n_QV^*Q$ where $n$ is the fiber dimension
of $Q\to\Si$, let us consider the corresponding metaplectic bundle
$\cD_Q\to Q$ and the tensor product $Y_Q=C_Q\op\ot_Q\cD_Q$,
where $C_Q=Q\times\bC$
is the trivial complex line bundle over $Q$. It is readily observed that
\be
\pi_{VQ}^*Y_Q=
C\op\ot_{V^*Q}\pi^*_{VQ}\cD_Q
\ee
and that the Hamiltonian vector fields
\be
\vt_f=a^k\dr_k-(p_r\dr_ka^r +\dr_kb)\dr^k, \qquad f=a^k(t,\si^\la,q^r)p_k
+ b(t,\si^\la,q^r),
\ee
of elements $f\in\cA_\cF$ are projectable onto $Q$. Then one can
associate to each
element $f$ of the quantum algebra $\cA_\cF$ the first order
differential operator
\mar{lmp135}\beq
\wh f=(-i\nabla_{\vt_f} +f)\ot\id+\id\ot\bL_{\pi_{VQ}(\vt_f)}=
-ia^k\dr_k-\frac{i}{2}\dr_ka^k+b \label{lmp135}
\eeq
in the space $\pi^*_{VQ}E_Q$ of sections of the pull-back bundle 
$\pi_{VQ}^*Y_Q\to
V^*Q$ which is
the pull-back of the space $E_Q$ of sections of $Y_Q\to Q$ whose
restriction to each fiber of $Q\to\Si$ is of compact support.
Since the pull-back $i^*_{Q_\si}\cD_Q$ of $\cD_Q$ onto each fiber $Q_\si$
of the fiber bundle $Q\to\Si$ is the metaplectic bundle over $Q_\si$,
the restrictions of elements of $\pi^*_{VQ}E_Q$ to a fiber $F=V^*_\si Q$
of
$V^*Q\to\Si$ constitute a pre-Hilbert space with
respect to the non-degenerate Hermitian form
\be
\lng i^*_F\pi_{VQ}^*\rho_Q|i^*_F\pi_{VQ}^*\rho'_Q\rng_\si=\op\int_{Q_\si}
i^*_{Q_\si}\rho_Q\ol{i^*_{Q_\si}\rho'_Q}.
\ee
The Schr\"odinger operators (\ref{lmp135}) are Hermitian operators in 
the pre-Hilbert
$C^\infty(\Si)$-module $\pi^*_{VQ}E_Q$, and
provide a desired leafwise geometric
quantization of the symplectic foliation $(V^*Q\to\Si,\Om_\cF)$.
Furthermore, one can replace this representation with the equivalent
representation by operators (\ref{lmp135}) in the pre-Hilbert module $E_Q$ of
sections of $Y_Q\to Q$.

\bigskip

\noindent
{\bf VI. CLASSICAL EVOLUTION EQUATION}
\bigskip

In order to quantize the evolution equation of a time-dependent
mechanical system, one should bear in mind that this equation is not reduced to
the Poisson bracket on $V^*Q$, but
is expressed into the Poisson bracket $\{,\}_T$ on the cotangent bundle
$T^*Q$ of $Q$.$^{14,21}$

Let us start from Hamiltonian dynamics of a classical mechanical
system with parameters on a configuration space $Q$ (\ref{pr1}). It is
convenient to assume for a time that
parameters are dynamic variables. The momentum phase space of such a
system is the vertical cotangent bundle $V^*_RQ$ of the configuration bundle
$Q\to\bR$ provided with holonomic coordinates
$(t,\si^\la,q^k,p_\la,p_k)$.$^{8,9}$
A Hamiltonian on this momentum phase space is defined
as a global section
\mar{qq4}\beq
h:V^*_RQ\to T^*Q, \qquad p\circ
h=-\cH(t,\si^\la,q^k,p_\la,p_k), \label{qq4}
\eeq
of the affine bundle
\be
\zeta_R:T^*Q\to V^*_RQ, \qquad (t,\si^\la,q^k,p,p_\la,p_k)\mapsto
(t,\si^\la,q^k,p_\la,p_k).
\ee
Given the canonical Liouville form $\Xi$ on $T^*Q$,
every Hamiltonian $h$ (\ref{qq4})
yields the pull-back Hamiltonian form
\mar{b4210}\beq
H=h^*\Xi= p_\la d\si^\la+ p_k dq^k -\cH dt  \label{b4210}
\eeq
on $V^*_RQ$.
For any Hamiltonian form $H$ (\ref{b4210}), there exists a unique
vector field $\g_H$ on $V^*_RQ$ such that
\be
\g_H\rfloor dt=1, \qquad \g_H\rfloor dH=0. \label{qq1}
\ee
This vector field defines the first order Hamilton equations on 
$V^*_RQ$.$^{7-9,14,21,22}$
Accordingly,
\mar{m59}\beq
\g_H\rfloor df=0, \qquad f\in C^\infty(V^*_RQ), \label{m59}
\eeq
is the evolution equation.
In order to express it into a
Poisson bracket, let us consider the pull-back
$\zeta^*_RH$
of the
Hamiltonian form
$H=h^*\Xi$ onto the cotangent bundle $T^*Q$. It is readily observed that the
difference
$\Xi-\zeta^*_Rh^*\Xi$ is a horizontal one-form on $T^*Q\to\bR$ and that
\mar{mm16}\beq
\cH^*=\dr_t\rfloor(\Xi-\zeta^* h^*\Xi))=p+\cH \label{mm16}
\eeq
is a function on $T^*Q$. Then the evolution equation
(\ref{m59}) is
brought into the form
\be
\{\cH^*,\zeta^*_Rf\}_T=0,
\ee
adapted for quantization.$^{14,21}$

Let us return to a system where $\si^\la$ are parameters.
Its Hamiltonian $\cH$ is affine in momenta $p_\la$, namely,
\mar{pr30}\beq
\cH=p_\la\G^\la +p_k(\La^k + \G^\la\La^k_\la) 
+\cH_\La(t,\si^\la,q^r,p_r), \label{pr30}
\eeq
where $\cH_\La\in
\zeta^*C^\infty(V^*Q)$ is the pull-back of a function on $V^*Q$ and
\be
\La\circ\G=dt\ot(\dr_t +\G^\la\dr_\la +(\La^k + \G^\la\La^k_\la)\dr_k)
\ee
is a connection on $Q\to\bR$ which is the composition of a connection
\mar{pr31}\beq
\G=dt\ot(\dr_t +\G^\la\dr_\la) \label{pr31}
\eeq
on the parameter bundle $\Si\to\bR$ and a connection
\mar{pr32}\beq
\La=dt\ot(\dr_t +\La^k\dr_k) +d\si^\la\ot(\dr_\la +\La^k_\la\dr_k) \label{pr32}
\eeq
on $Q\to\Si$.$^{7-9}$ Note that the second term of the connection
(\ref{pr32}) provides the lift
\be
\tau^\la\dr_\la\mapsto \tau^\la(\dr_\la +\La^k_\la\dr_k)
\ee
onto $Q$ of vertical vector fields on the parameter bundle $\Si\to\bR$. 
It plays the role of a control operator in holonomic quantum computation.
If a parameter function $\chi:\bR\to\Si$ is given, the 
connection
$\G$ (\ref{pr31}) on $\Si\to\bR$ is determined in such a way that
\mar{lmp55}\beq
\nabla^\G \chi=0,\qquad \G^\la(t,\chi^\m(t))=\dr_t\chi^\la. \label{lmp55}
\eeq

It is readily observed that, if a Hamiltonian $\cH$ is affine in
momenta $p_\la$ and if $f$ is a function on $V^*Q$, then the bracket
$\{\cH^*,\zeta^*f\}_T$, where $\zeta$ is the fibration (\ref{lmp44}),
is the pull-back of
a function on $V^*Q$. It provides a derivation of
the $\bR$-ring $C^\infty(V^*Q)$. Therefore,
one can think of the equality
\mar{pr35}\beq
\{\cH^*,\zeta^*f\}_T=0, \qquad f\in C^\infty(V^*Q),  \label{pr35}
\eeq
as being a classical evolution equation on $C^\infty(V^*Q)$.

\bigskip

\noindent
{\bf VII. QUANTUM EVOLUTION EQUATION}
\bigskip

In order to quantize the evolution equation (\ref{pr35}), one should 
quantize the
Poisson manifold $(T^*Q,\{,\}_T)$ so that its quantum algebra
$\cA_T$ contains $\zeta^*\cA_\cF$. Let $\Phi$ be polarization of the
Poisson manifold $(V^*Q,\{,\}_V)$ which determines $\cA_\cF$.
Then, by virtue of the relation (\ref{m72'}), $\zeta^*\Phi$ is a polarization
of $(T^*Q,\{,\}_T)$. Clearly, $\cA_\cF$ is a subalgebra
of the quantum algebra $\cA_T$ of $T^*Q$ determined by this polarization.
The quantum algebra $\cA_T$ consists of functions on $T^*Q$ which are
affine in momenta $p,p_\la,p_k$. Let us restrict our consideration to its
subalgebra $\cA'_T$ of functions
\be
f=a(t,\si^\m)p +a^\la(t,\si^\m)p_\la + a^k(t,\si^\m,q^r)p_k + b(t,\si^\m,q^r),
\ee
where $a$ and $a^\la$ are the pull-back onto $T^*Q$ of functions on
the parameter space $\Si$. Using transformation laws of momenta $p$, 
$p_\la$ and $p_k$, one can
justify that this notion is coordinate-independent.
Of course, $\cA_\cF\subset \cA'_T$.
Moreover, $\cA'_T$ admits a representation by the Hermitian operators
\be
\wh f= -i(a\dr_t + a^\la\dr_\la +a^k\dr_k)-\frac{i}{2}\dr_ka^k+b
\ee
in the same pre-Hilbert $C^\infty(\Si)$-module $E_Q$ as the representation
(\ref{lmp135}) of $\cA_\cF$.
Then, if $\cH^*\in \cA'_T$, the evolution equation (\ref{pr35}) is quantized
as the Heisenberg equation
\mar{lmp50}\beq
i[\wh \cH^*,\wh f]=0, \qquad f\in \cA_\cF. \label{lmp50}
\eeq

The problem is
that the function $\cH^*$ (\ref{mm16}) fails to
belong to the
algebra $\cA'_T$, unless the Hamiltonian function $\cH_\La$ 
(\ref{pr30}) is affine in
momenta $p_k$. Let us assume that $\cH_\La$ is polynomial in momenta.
This is the case of almost all physically relevant models.

\begin{lem} \label{lmp70} \mar{lmp70}
Any smooth function $f$ on $V^*Q$ which is a polynomial of momenta
$p_k$ is decomposed in
a finite sum of products of elements of the algebra $\cA_\cF$.
\end{lem}

\begin{proof}
A polynomial $f$ is
a sum of homogeneous polynomials of fixed degree in momenta.
Therefore, it suffices to prove the statement for
an arbitrary homogeneous polynomial $f$ of degree $n>1$
on $V^*Q$. We use the fact that the vertical cotangent bundle 
$V^*Q\to Q$ admits a
finite bundle atlas.$^{23}$ Let $\{U_\xi\}$, $\xi=1,\ldots,r$,
be the corresponding open cover of
$Q$ and $\{\vf_\xi\}$ a smooth partition of unity subordinate to this cover.
Put
\be
l_\xi=\vf_\xi(\vf_1^n+\cdots +\vf_r^n)^{-1/n}.
\ee
It is readily observed that $\{l_\xi^n\}$ is also a partition of unity
subordinate to $\{U_\xi\}$. Let us consider the polynomials
\be
f_\xi=f|_{U_\xi}=\op\sum_{(k_1\ldots k_n)} a^{k_1\ldots
k_n}_\xi(q)p_{k_1}\cdots p_{k_n},
\qquad q	\in U_\xi.
\ee
Then we obtain a desired decomposition
\mar{ooo}\beq
f=\op\sum_\xi l^n_\xi f_\xi=\op\sum_\xi\op\sum_{(k_1\ldots k_n)}
   [l_\xi a^{k_1\ldots k_n}_\xi p_{k_1}][l_\xi p_{k_2}]
\cdots [l_\xi p_{k_n}],
\label{ooo}
\eeq
where all terms $l_\xi a^{k_1\ldots k_n}_\xi p_{k_1}$ and $l_\xi p_{k_i}$
are smooth functions on $V^*Q$.
\end{proof}

By virtue of Lemma \ref{lmp70}, one can associate to a polynomial
Hamiltonian function $\cH_\La$ an element of the enveloping
algebra $\ol\cA_\cF$ of the Lie algebra $\cA_\cF$. Accordingly,
$\cH^*$ is represented by an element of the enveloping algebra
$\ol\cA'_T$ of the Lie algebra $\cA'_T$.
Then the Schr\"odinger representation of
the Lie algebras $\cA'_T$ and
$\cA_\cF$ is naturally extended to their enveloping algebras $\ol\cA'_T$
and $\ol\cA_\cF$ that provides quantization of $\cH^*$.

Of course, the decomposition (\ref{ooo})
by no means is unique.
An ambiguity of the operator representation of a classical
Hamiltonian that  does not
preserve a polarization is
a well-known technical problem of Schr\"odinger quantization as like as
of any geometric quantization scheme.$^{13,14}$

Given an operator $\wh\cH^*$, the bracket
\mar{qq120}\beq
\nabla\wh f= i[\wh \cH^*,\wh f] \label{qq120}
\eeq
defines a derivation of the quantum algebra $\ol\cA_\cF$.
Since $\wh
p=-i\dr_t$, the derivation (\ref{qq120}) obeys the Leibniz rule
\be
\nabla (r\wh f)=\dr_t r\wh f + r\nabla \wh f, \qquad r\in
C^\infty(\bR).
\ee
Therefore, it is a connection on the $C^\infty(\bR)$-algebra 
$\ol\cA_\cF$, which
enables one to treat quantum evolution of $\ol\cA_\cF$ as a parallel transport
along time.$^{7,8,14}$
In particular, $\wh f$ is parallel with respect to
the connection (\ref{qq120}) if it obeys the Heisenberg equation (\ref{lmp50}).
Given a trivialization
\mar{lmp71}\beq
Q\cong \bR\times M \label{lmp71}
\eeq
and the corresponding (global) decomposition
\be
\wh\cH^*=-i\dr_t +\wh\cH,
\ee
we can introduce the evolution operator $U$
which obeys the equation
\be
\wh\cH^*\circ U=-iU\circ\dr_t,
\ee
and  can be written as the formal time-ordered exponent
\be
U=T\exp\left[-i\op\int^t_0\wh\cH dt'\right].
\ee
One can think of $U$ as being an operator of the
parallel displacement in the $C^\infty(\bR)$-module
$E_Q$ with respect to the connection
\be
\nabla\rho_Q=i\wh\cH^*\rho_Q, \qquad \rho_Q\in E_Q.
\ee

Now let us consider a mechanical system depending on a given parameter
function $\chi:\bR\to\Si$.
Its configuration space is the pull-back bundle $Q_\chi=\chi^*Q$
over $\bR$ which is a subbundle
$i_\chi:Q_\chi\to Q$ of the fiber bundle $Q\to\bR$. The corresponding
momentum phase space is the pull-back bundle $i^*_\chi V^*Q$, isomorphic
to the vertical cotangent bundle $V^*Q_\chi$ of $Q_\chi\to\bR$. The 
pull-back of the
Hamiltonian form $H$ (\ref{b4210}) onto $V^*Q_\chi$, where the 
connection $\G$ obeys the
relation (\ref{lmp55}), reads
\be
&& H_\chi=p_kdq^k -\cH_\chi dt, \\
&& \cH_\chi=p_k(\La^k(t,\chi^\m(t),q^r)+
\La^k_\la(t,\chi^\m(t),q^r)\dr_t \chi^\la) +
\cH_\La(t,\chi^\m(t),q^r,p_r).
\ee
It characterizes the dynamics of a mechanical system with a given parameter
function $\chi$.$^{7-9}$

In order to quantize this system, let us consider the pull-back
bundle $\cD_\chi=i_\chi^*\cD_Q$ over $Q_\chi$
and its pull-back sections $\rho_\chi=i_\chi^*\rho_Q$, $\rho_Q\in 
E_Q$. It is easily
justified that these are
leafwise half-forms on the fiber bundle $Q_\chi\to\bR$ whose restrictions
to each fiber $i_t:Q_t\to Q_\chi$ are of compact support. These 
sections constitute
a pre-Hilbert $C^\infty(\bR)$-module $E_\chi$ with respect to the 
Hermitian forms
\be
\lng i^*_t\rho_\chi|i^*_t\rho'_\chi\rng_t=\op\int_{Q_t}
i^*_t\rho_\chi\ol{i^*_t\rho'_\chi}.
\ee
Then the pull-back operators
\be
&& (\chi^*\wh f)\rho_\chi=(\wh f\rho)_\chi, \\
&& \chi^*\wh f=
  -ia^k(t,\chi^\la(t),q^r)\dr_k
-\frac{i}{2}\dr_ka^k(t,\chi^\m(t),q^r) + b(t,\chi^\m(t),q^r),
\ee
in $E_\chi$ provide the representation of the pull-back functions
\be
i_\chi^*f=a^k(t,\chi^\la(t),q^r)p_k + b(t,\chi^\la(t),q^r), \qquad 
f\in \cA_\cF,
\ee
on $V^*Q_\chi$. Accordingly, the quantum operator
$\wh\cH_\chi^*=\wh p+\wh \cH_\chi$ coincides with pull-back operator
$\chi^*\wh\cH^*$.
Then the Heisenberg equation of a quantum system with a parameter function
$\chi$ takes the form
\be
i[\wh\cH^*_\chi,\chi^*\wh f]=0,
\ee
and the corresponding evolution operator reads
\mar{lmp60}\beq
U=T\exp\left[-i\op\int^t_0 \wh\cH_\chi dt'\right]. \label{lmp60}
\eeq

The key point is that the Hamiltonian $\wh\cH_\chi$ in the evolution
operator $U$ (\ref{lmp60}) is affine in the derivatives 
$\dr_t\chi^\la$, namely,
\mar{lmp72}\beq
\wh\cH_\chi=-i(\La^k_\la\dr_k +\frac12 \dr_k\La^k_\la)\dr_t\chi^\la + \wh \cH'.
\label{lmp72}
\eeq
Its first term
is responsible for the geometric Berry factor phenomena as follows,
while $\wh\cH'$ can be
regarded as a dynamic Hamiltonian of a quantum system.

Bearing in mind possible applications to holonomic quantum
computations, let us simplify the quantum system in question.
Given a trivialization (\ref{lmp71}) of a configuration bundle $Q$, we have the
corresponding trivialization of the parameter bundle $\Si=\bR\times 
S$ such that
the fibration $Q\to\Si$ reads
\be
\bR\times M\ar^{\id\times \pi_M}\bR\times S,
\ee
where $\pi_M: M\to S$ is a fiber bundle. Note that, from the physical 
viewpoint,
a trivialization (\ref{lmp71}) provides a reference frame in nonrelativistic
mechanics.$^{9,22}$
Let us suppose that there exists a reference frame such that the
components $\La^k_\la$ of the connection $\La$ (\ref{pr32}) are independent
of time. Then one can regard the second term in this connection as a connection
on the fiber bundle $M\to S$. It also follows that the first term
in the Hamiltonian (\ref{lmp72}) depends on time
only through parameter functions $\chi^\la(t)$. Furthermore, let the two terms
in the  Hamiltonian (\ref{lmp72})
mutually commute on $[0,t]$.  Then the evolution
operator
$U$ (\ref{lmp60}) takes the form
\be
U=
T\exp\left[ -\op\int_{\chi([0,t])}(\La^k_\la\dr_k 
+\frac12\dr_k\La^k_\la) d\si^\la \right]
\circ T\exp\left[ -i\op\int_0^t\wh\cH' dt'\right].
\ee
One can think of the first factor in this evolution operator as being the
parallel displacement operator
along the curve
$\chi([0,t])\subset S$ with respect to the connection
\be
\nabla^\La\rho_Q=(\dr_\la + \La^k_\la\dr_k +\frac12\dr_k\La^k_\la 
)\rho_Q d\si^\la,
\qquad \rho_Q\in E_Q,
\ee
on the $C^\infty(S)$-module $E_Q$.
Its peculiarity in comparison with the remaining
one lies in the fact that integration over
time through a parameter function $\chi(t)$ depends only
on a trajectory of this function in a parameter space, but not
on parametrization of this trajectory by time. Therefore,
one can think of it as being a geometric factor.


\begin{thebibliography}{ederf}




\bibitem{zan} P.Zanardi and M.Rasetti, {\it Phys. Lett.} {\bf A264}, 94 (1999).

\bibitem{fuj} K.Fujii, {\it J. Math. Phys.} {\bf 41}, 4406 (2000).

\bibitem{pach} J.Pachos and P.Zanardi, {\it Int. J. Mod. Phys.} {\bf B15},
1257 (2001).

\bibitem{wilcz} F.Wilczek and A.Zee, {\it Phys. Rev. Lett.} {\bf 52},
2111 (1984).

\bibitem{anan} J.Anandan and Y.Aharonov, {\it Phys. Rev} {\bf D38},
1863 (1988). 

\bibitem{bohm} A.Bohm and A.Mostfazadeh, {\it J. Math. Phys.}
{\bf 35}, 1463 (1994).

\bibitem{sard00} G.Sardanashvily, {\it J. Math. Phys.} {\bf 41},
5245 (2000).

\bibitem{book00} L.Mangiarotti and G.Sardanashvily, {\it Connections in
Classical and Quantum Field Theory} (World Scientific, Singapore, 2000).

\bibitem{book98} L.Mangiarotti and G.Sardanashvily, {\it Gauge Mechanics}
(World Scientific, Singapore, 1998).

\bibitem{vais91} I.Vaisman, {\it J. Math. Phys.} {\bf 32}, 3339
(1991).

\bibitem{vais} I.Vaisman, {\it Lectures on the Geometry of Poisson
Manifolds} (Birkh\"auser Verlag, Basel, 1994).

\bibitem{dixm2} J.Dixmier, {\it Alg\`ebras Enveloppantes}
(Gauthier-Villars \"editeur, Paris, 1974).

\bibitem{sni} J.\'Sniatycki, {\it Geometric Quantization and Quantum
Mechanics} (Springer-Verlag, Berlin, 1980)

\bibitem{jmp01} G.Giachetta, L.Mangiarotti and G.Sardanashvily, {\it J.
Math. Phys.} {\bf 43}, 56 (2002).

\bibitem{vais97} I.Vaisman,  {\it Diff. Geom. Appl.} {\bf 7}, 265 (1997).


\bibitem{epr} G.Sardanashvily, E-print arXiv: math.DG/0110196.

\bibitem{hect} G.Hector, E.Mac\'ias  and M.Saralegi,
{\it Publ. Mat.} {\bf 33}, 423 (1989).

\bibitem{vais73} I.Vaisman, {\it Cohomology and Differential Forms}
(Marcel Dekker, Inc., New York, 1973).

\bibitem{wood} N.Woodhouse, {\it Geometric Quantization} (Clarendon
Press, Oxford, 1992).

\bibitem{eche98} A.Echeverr\'{\i}a Enr\'{\i}quez, M.Mu\~noz Lecanda,
N.Rom\'an Roy and C.Victoria-Monge, {\it Extracta Math.} {\bf 13},
135 (1998).

\bibitem{jmp00} L.Mangiarotti and G.Sardanashvily, {\it J. Math. Phys.}
{\bf 41}, 2858 (2000).

\bibitem{sard98} G.Sardanashvily,
{\it J. Math. Phys.} {\bf 39},  2714 (1998)

\bibitem{greub} W.Greub, S.Halperin and R.Vanstone, {\it Connections,
Curvature, and Cohomology}, Vol. 1,  (Academic Press, N.Y., 1972).

\end{thebibliography}
\end{document}